# Mid-Infrared Ultrafast Carrier Dynamics in Thin Film Black Phosphorus

*Vasudevan Iyer[1], Peide Ye[2], and Xianfan Xu[1*]*

[1]Department of Mechanical Engineering and Birck Nanotechnology Center, Purdue University, West Lafayette, 47907, USA

[2]Department of Electrical and Computer Engineering and Birck Nanotechnology Center, Purdue University, West Lafayette, 47907, USA

Black phosphorus is emerging as a promising semiconductor for electronic and optoelectronic applications. To study fundamental carrier properties, we performed ultrafast femtosecond pump-probe spectroscopy on thin film black phosphorus mechanically exfoliated on a glass substrate. Carriers (electrons and holes) were excited to high energy levels and the process of carrier relaxation through phonon emission and recombination was probed. We used a wide range of probing wavelengths up to and across the band gap to study the evolution of the relaxation dynamics at different energy levels. Our experiments revealed a plethora of important physical phenomena. The fast relaxation time constants, associated with carrier-phonon scattering, steadily increase as the energy of the probe beam approaches the band gap energy, which was determined to be 0.31 eV, and the carrier recombination rate was obtained when the probe wavelength was

Email correspondence to xxu@ecn.purdue.edu

tuned to match the band gap energy. The carrier-phonon scattering rates were found to be similar along the armchair and zigzag directions, therefore, the anisotropic carrier mobility reported in literature is mainly due to the difference in effective mass of carriers along different directions. The ultrafast spectroscopy data further revealed the oxidation induced surface charges. Our results highlight the importance of using the spectroscopy technique, in this case, in the mid-IR range, to uncover useful physical processes.

**Keywords**: Black Phosphorus, Mid-Infrared, Carrier dynamics, Band edge, Ultrafast Spectroscopy

Since the last decade there has been an increasing interest in the study of two-dimensional materials, especially graphene and transition metal dichalcogenides (TMDC) for fabricating many electronic and optoelectronic devices [1-11]. Black phosphorus has emerged as a recent material in this field and has shown promises with its direct bandgap, anisotropic electronic, optical, and thermal properties, and good device performance [12-19]. Black phosphorus has a direct bandgap in the bulk of around 0.35 eV and can become as large as 1.5 eV in the monolayer form according to first principles calculations, photoluminescence studies, and electrical measurements [13, 20-24]. The direct bandgap is attractive for use in electronic devices such as solar cells [25, 26] and photodetectors [27-30]. Black phosphorus has also been successfully tested as a transistor channel material [31-33].

Given the vast interest in black phosphorus, it is important to investigate its properties at a fundamental level. Ultrafast optical spectroscopy is very useful in understanding carrier scattering, relaxation, and recombination processes. Many ultrafast studies have been carried out on graphene [34-40] and TMDC such as $MoS_2$ [41-44]. There have also been a few ultrafast optical studies on black phosphorus which have brought out the enhancement of anisotropic behavior after photoexcitation [45], provided upper bounds for mobility values through spatial scanning study [46], and highlighted the carrier decay process at long time scales which was attributed to carrier lifetime, i.e. recombination [47]. A recent article investigated the decay process of excited carriers at short time scales on liquid exfoliated black phosphorus nanosheets and determined the carrier-phonon scattering times at a few different probing energy levels corresponding to commonly encountered laser emission lines [48].

Most ultrafast pump-probe studies are performed with one excitation wavelength and one probe wavelength. On the other hand, probing a range of wavelengths or equivalently a range of energy states, especially near the band edge, can uncover the vital information about the carrier relaxation process as demonstrated for graphene and $MoS_2$ [34, 37, 49]. In this study, we focus on understanding the evolution of the decay of excited states with varied probe energy approaching the band gap of black phosphorus, and relate the ultrafast dynamics with fundamental transport properties. The scattering processes in black phosphorus in different directions are studied in order to gauge their importance in the observed anisotropic carrier mobility. A previous study[48] obtained a direction averaged scattering rate using randomly oriented flakes produced by liquid exfoliation and present the carrier dynamics at select probing energies. We systematically varied the probing direction and energies on a mechanically exfoliated flake, and find carrier scattering rates along the armchair and zigzag directions with a saturation of scattering time as the probe energy is tuned closer to and below the band edge, which helps to reveal the origin of anisotropic transport properties. The information on recombination was obtained when the probing wavelength was tuned to the band gap energy. Finally, the ultrafast spectroscopy data also revealed surface charge induced by oxidation.

**Results and Discussion**

The transient reflectance data obtained using the 792 nm pump wavelength polarized along the armchair direction and probe wavelengths varied from 1,700 nm to 2,200 nm, also polarized along the armchair direction are presented in Figure 1a. Data for wavelengths from 2,500 nm to 4,600 nm are shown in Figure 1b. All the datasets have been normalized by their minimum values to -1. Change in reflectance is defined as the reflectance after pump excitation minus the reflectance without any pump excitation.

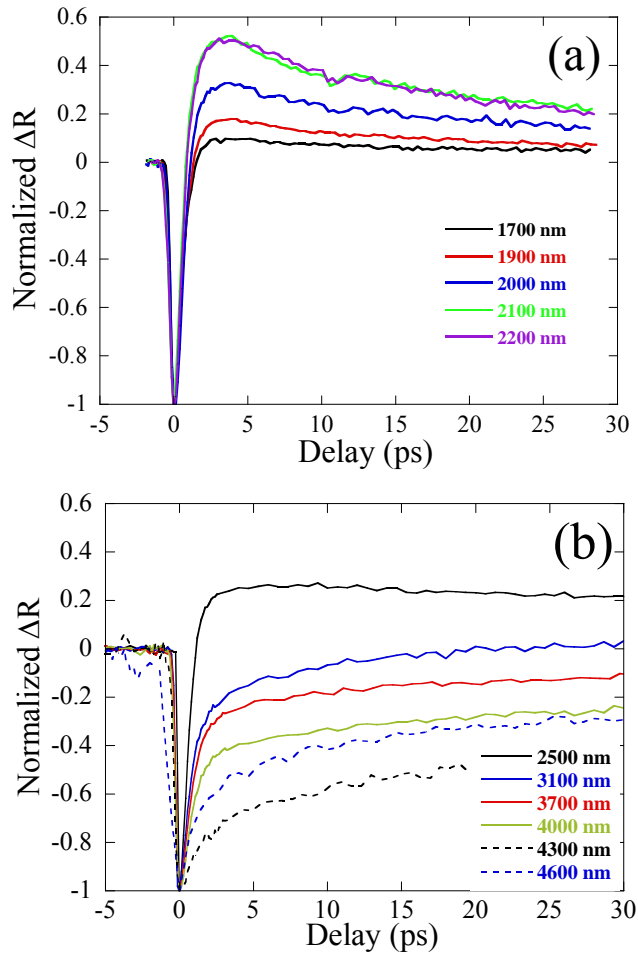

Figure 1: Transient reflectivity of black phosphorus for probing wavelengths from (a) 1,700 nm to 2,200 nm and (b) 2,500 nm to 4,600 nm.

During photoexcitation by a femtosecond laser pulse, carriers are generated at those energy levels in the conduction and valence band that correspond to vertical transitions in an E-k band diagram at the pumping energy. These carriers have momentum along the polarization direction of the pump. They then randomize their momentum and energy by carrier-carrier scattering processes. This randomization process is fast and beyond the measurement time resolution (See supporting information 2). The next step is the formation of a thermalized distribution which can be represented by the Fermi-Dirac distribution. In order to achieve thermalization, carriers must move closer to the band edge in energy space and this implies intraband relaxation. The first decay in the experimental data starting at zero picosecond (rising from $\Delta R = -1$) and of the order of one picosecond is attributed to Pauli blocking and carrier-phonon scattering leading to the formation of a thermalized carrier distribution [47, 48]. Pauli blocking is a result of occupation of carriers at the probing energy level. As the carriers emit phonons and relax to states closer to the band edge, the contribution from Pauli blocking diminishes, reflecting in the rise in the reflectivity signal. The second decay that starts from a few picoseconds, and is much slower, is attributed to the carrier recombination and lattice heating, and the time scales (10-100 picoseconds) are in good agreement with published work [45-47]. This second decay is thus an interband relaxation process. As can be seen from Figure 2, for wavelengths closer to the band edge, $\Delta R_{max}/R_0$ is greater. This is because energies closer to the band edge have a higher carrier population after thermalization and thus produce a higher $\Delta R$. The data shown in Figure 2 also indicates that the bulk bandgap is close to 4,000 nm (0.31 eV), in agreement with previous theoretical work[13, 20, 24], recent FTIR data[50], and electrical measurements[23]. Additionally, a second peak about 40 meV below the band edge was also observed, which can be attributed to either exciton as theoretically predicted[20] or acceptor energy state as experimentally observed[51, 52]. Supporting information 3 provides transmission data taken for a cluster of randomly oriented flakes. The band gap and sub-band gap excitation energy obtained from the transmission measurements fully agree with the ultrafast spectroscopy data presented in Figure 2. The transient signal vanishes when the probe wavelength is above 4,600 nm (below the sub-band gap energy state, supporting information 4).

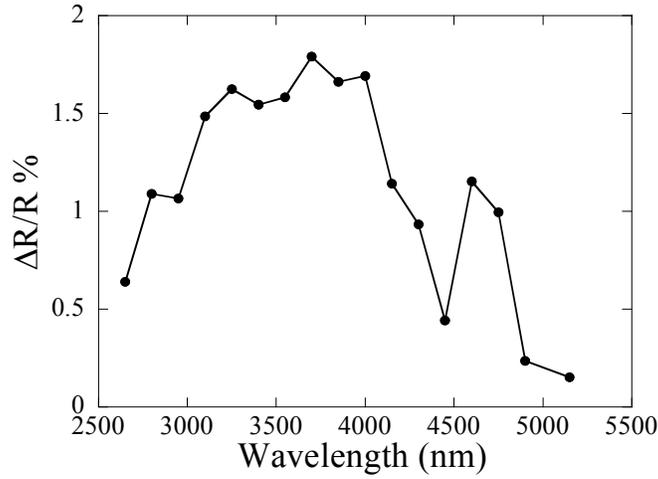

Figure 2: ΔR(t=0)/R0 vs wavelength showing the bandgap near 0.31 eV and the presence of sub-band gap energy state.

Bi-exponential fitting is used to extract the time constants for the fast and slow decays at different wavelengths:

$$\Delta R_{t>0} = A * \exp\left(-\frac{t}{\tau_1}\right) + B * \exp\left(-\frac{t}{\tau_2}\right) + C \quad (1)$$

The constant term $C$ is attributed to the much slower thermal energy dissipation due to lattice heating. $A$ and $B$ are the amplitudes for the exponentials. $\tau_1$ and $\tau_2$ are the fast and slow decay times. Examples of fitting for the 2,200 and 4,000 nm probes using Equation (1) are shown in supporting information 5, which show that the bi-exponential function can represent the signals well.

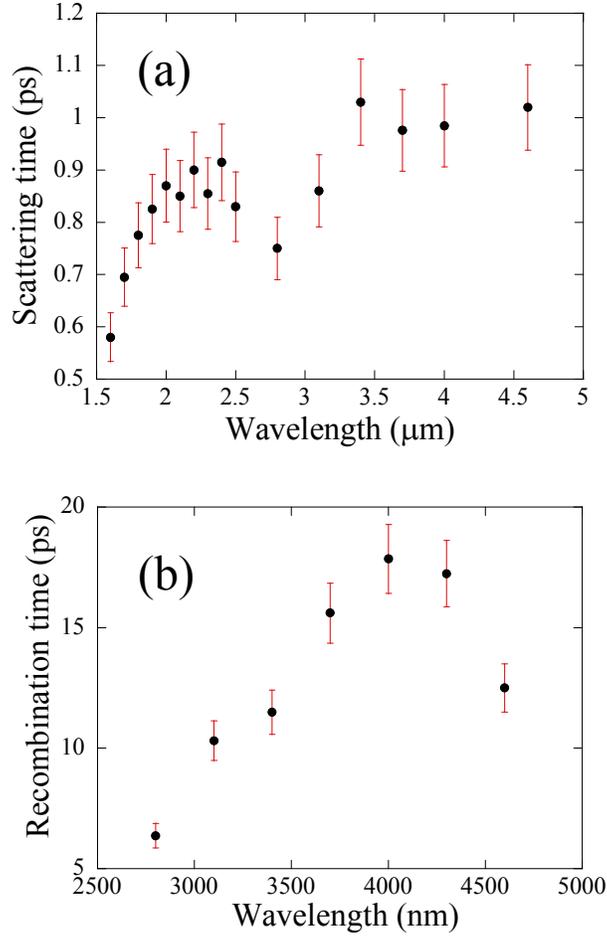

Figure 3: (a) Fast and (b) slow time constants with different probe energy. The fit uncertainty is about 8%.

Figure 3a shows that the carrier-phonon scattering time at different energy levels approaches a value of about 1 ps as the probe energy approaches the band gap energy. This increase in the scattering time can be understood by the fact that as carriers approach the band edge, there are fewer channels to dissipate energy in the intraband relaxation process. From the Raman measurements, the optical phonon energy for black phosphorus is about 50 meV. Probing the sample with 0.4 eV (3,100 nm) photons corresponds approximately to 50 meV above the band edge in both conduction and valence band. Since the excited carriers are expected to form a thermalized distribution a few meV above the band edge, carriers probed by 0.4 eV light are at the limit of one optical phonon emission, and probing with photon energy less than 0.4 eV leads to the relaxation time saturation behavior.

Using the carrier-phonon scattering times $\tau$ obtained at the probing wavelength corresponding to the band gap, and $\mu = \frac{q\tau}{m^*}$, carrier mobility can be calculated. It is known that black phosphorus exhibits anisotropic transport properties along its armchair and zigzag directions such as mobility [12, 13]. Therefore, carrier-phonon scattering was also measured along the zigzag direction and is shown in the supporting information 6. An ambipolar mobility of about 1900 cm²/V·s along the zigzag direction and 22,000 cm²/V·s along the armchair direction are obtained. The effective mass of electrons (holes) along the armchair direction is taken as 0.0825 (0.0761) $m_0$ whereas along the zigzag direction is 1.027 (0.648) $m_0$ [53]. Since electrons and holes have similar effective mass, the ambipolar mobility, defined as $\mu_a = \frac{2\mu_e\mu_h}{\mu_e + \mu_h}$, can be considered as an average carrier mobility. The ambipolar mobility is determined since the experimental data is a result of relaxation of both electrons and holes. The scattering time obtained is the upper limit for both holes and electrons, and given the comparable valence and conduction bands, similar scattering times are expected for electrons and holes. Our estimation of the mobility is reasonable compared to previous ultrafast experimental work (but with a probe wavelength of 810 nm) which determined a value of around 3,000 cm²/V·s for the zigzag direction and 50,000 cm²/V·s for the armchair direction [46]. From Figure 3 we can see the importance of probing near the bandgap to extract the correct scattering rates needed in calculations. Theoretical study predicted values in the $10^3$ cm²/V·s range [12]. However for fabricated devices, the mobilities were found not to exceed 1,000 cm²/V·s [13, 16, 31, 33, [54, 55]. The bulk and surface impurity scattering at equilibrium, which affects only the momentum of carriers (not their energy), would account for further reduction in mobility in devices which were not captured in the ultrafast spectroscopy experiments that are sensitive to changes in energy. Thus our experiment provides an upper limit for mobility.

The slow decay time (recombination time), $\tau_2$, also increases when the probing wavelength approaches the band gap energy. This is because the carriers, not at the band edge (4,000 nm), scatter with phonons and move toward low energy levels. Hence, due to the additional relaxation channel, the higher energy carriers appear to have a faster recombination time. The recombination time, independent of carrier phonon scattering, must be evaluated at the bandgap energy. At the 4,000 nm probing wavelength, the recombination time $\tau_2$ is found to be about 18 ps. Figure 3b also shows that the recombination time at sub-band gap wavelength 4,600 nm is about 12 ps. Since the data indicate both sub-band gap carriers and carriers in the delocalized bands relax with a comparable time, we expect a similar recombination mechanism for both, e.g., phonon assisted recombination. It is likely that the sub-band gap signal arises from acceptor states (vs. from excitons) because it displays features similar to the band edge probe (4,000 nm). These states are filled after rapid carrier-carrier scattering (fast drop in ΔR) and subsequently carrier-phonon scattering displaces carriers from the sub-band gap states to the delocalized states (fast rise in ΔR). For excitons, we would expect a different dynamics because the bound states should possibly show much longer decay times[49, 56].

It is worthwhile to look more closely at the relaxation process as the wavelength is varied. For shorter wavelengths (e.g., 2,200 nm, Figure 1a), a positive peak is observed, but the peak does not rise above zero for wavelengths longer than 3,100 nm (Figure 1b). We propose that such optical response is related to surface charges. It is known that oxidation can induce surface doping on the order of $10^{12}$ cm$^{-2}$.[33] Photoexcited holes in the bulk thermalize close to the band edge whereas photoexcited holes at the surface thermalize with the existing hole population due to doping by oxidation (See supporting information 7). This creates an excess hole population at slightly higher energies in the valence band and changes the dielectric constant. The change in dielectric constant is reflected in the spectroscopy data, which indicates variation of the (peak) reflectivity at different probing wavelengths corresponding to different energy levels in the valence band. We performed a calculation of the change in the dielectric constant at the surface and bulk and related it to the change in reflectivity in supporting information 7. Results of the calculated reflectivity reproduces the observed positive change in reflectivity signals, as shown in Figure 4.

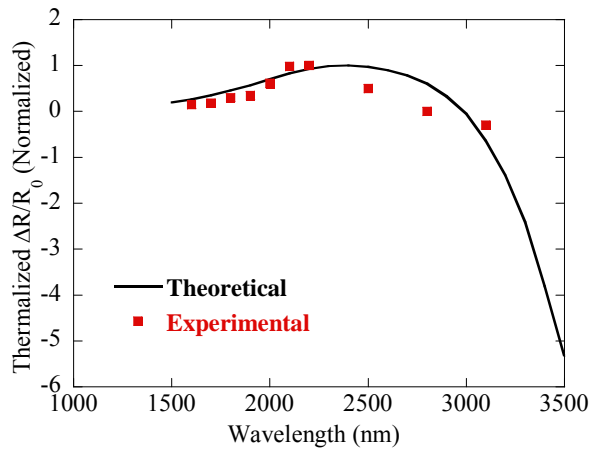

Figure 4: Theoretical estimate of ΔR/R and experimental values around 2,000 nm. The experimental points are at 3 ps delay, i.e. after thermalization.

**Conclusion**

In conclusion, we have carried out ultrafast infrared spectroscopy on thin film black phosphorus and found that the near band gap phonon-scattering time is close to 1 picosecond which is limited by the optical phonon scattering processes. By investigating energy relaxation of carriers with a tunable probe energy, it is found that carrier-phonon scattering time saturates when the probe energy reaches close to the band edge, which can be used to extract fundamental properties such as mobility. From the transient reflectivity data for different excitation and probing polarizations corresponding to the different anisotropic directions in black phosphorus, we find a similar scattering time along the armchair and zigzag directions. The major factor controlling the carrier mobility is thus the difference in the effective mass resulted from anisotropic band structure of the material rather than the scattering processes. The recombination time was also seen to increase as the probing wavelength approached the band edge. A sub-band gap energy state was also observed and attributed to acceptor states. Lastly, we used an optical property model for qualitatively explaining the spectroscopic features in the data, and found contributions due to surface doping by oxidation. Our work shows the importance of selecting the right probing wavelengths to study carrier dynamics in semiconductors displaying variety of interesting phenomena.

**Experimental Method**

A Coherent ultrafast Ti:Sapphire amplified laser system is used to generate 40 fs pulses at 792 nm center wavelength and 5 kHz repetition rate. The beam is split by a beam splitter and majority of the power is used to pump an optical parametric amplifier (OPA, Light Conversion) which generates wavelengths from 1.2 to 20 μm with 120 fs pulse duration. The weaker half is sent through a translation stage and is used to excite the sample. The OPA output is used as the probe. Once the pump and probe are made collinear, they are focused onto the sample surface using a 20x-0.42NA objective (Mitutoyo) for wavelengths up to 2.2 μm and with a parabolic mirror for longer wavelengths. The reflected light is re-collimated by the same objective/mirror and enters the photodetector (Thorlabs DET10D/Electro-optical systems MCT detector). The pump is chopped at a frequency of 500 Hz and lock-in detection is used to obtain good signal-to-noise ratio.

A half-waveplate is used to control the polarization of the pump, whereas the sample is rotated to align it parallel or perpendicular to the probe polarization. The diameter of the pump and probe spot is 10 μm and the probe is at least two orders of magnitude weaker than the pump. The pump energy is about 6 nJ per pulse and fluence dependence test confirmed a linear response of the acquired data around this pump energy (See supporting information 1). Pure nitrogen is blown over the sample to prevent oxygen and water vapor from reaching the surface and degrading the sample. All experiments were done at room temperature.

Bulk black phosphorus crystals are mechanically cleaved a few times on scotch tape to obtain thin film flakes with flat and freshly exposed surface. The flakes are then transferred to a clean glass substrate. To determine the arm-chair and zigzag directions of the flake, polarized Raman spectroscopy is performed, which is a fast and accurate method [1]. The Raman spectrum when the polarization direction is aligned with the armchair direction is shown in Figure 5, in agreement with previous studies. The $B_g^2$ peak (near 440 cm$^{-1}$) is nearly invisible, indicating good alignment between the laser polarization direction and the armchair direction.

The optical image of the sample is shown in Figure 5 as an inset. No degradation of the sample is observed before and after the experiments. This also is confirmed by the fact that the Raman signal, transient signal and appearance of the sample remained unchanged. The thickness of the sample is 65 nm as determined by atomic force microscopy (AFM) measurements.

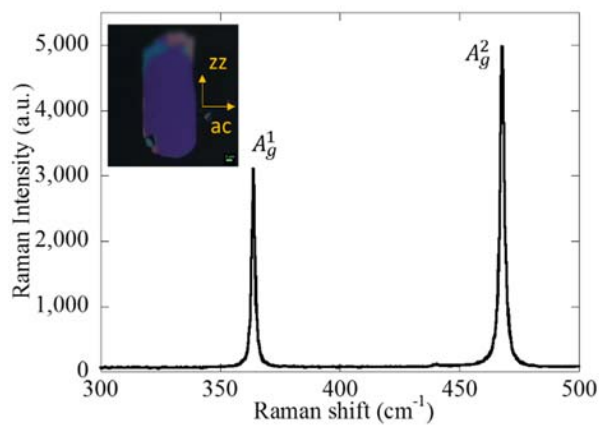

Figure 5: Armchair polarized Raman signal. Inset shows flake. Scale bar is 2 μm. zz stands for zigzag direction and ac stands for armchair direction.

**Acknowledgement**

The authors would like to thank Anurup Dutta for his help with the atomic force microscope measurement and Prabhu Kumar for dielectric constant measurement. This work was supported by AFOSR/NSF under EFRI 2-DARE Grant EFMA-1433459.

**Notes**

The authors declare no competing financial interest.

**Supporting Information**

Pump fluence dependence; Randomization of carriers in k-space; Transmittance data; Biexponential fitting; anisotropic probing; surface doping effects on transient signals

# Supporting Information

# Mid-Infrared Ultrafast Carrier Dynamics in Thin Film Black Phosphorus

*Vasudevan Iyer[1], Peide Ye[2], and Xianfan Xu[1]\**

[1]Department of Mechanical Engineering and Birck Nanotechnology Center, Purdue University, West Lafayette, 47907, USA

[2]Department of Electrical and Computer Engineering and Birck Nanotechnology Center, Purdue University, West Lafayette, 47907, USA

Email correspondence to xxu@ecn.purdue.edu

**Supporting Information 1 | Laser Fluence dependence**

We varied the laser power to confirm a linear response of the transient reflectance signals. The peak value of the signal was monitored for various pump powers at a probing wavelength of 2,000 nm as shown in Figure S1.

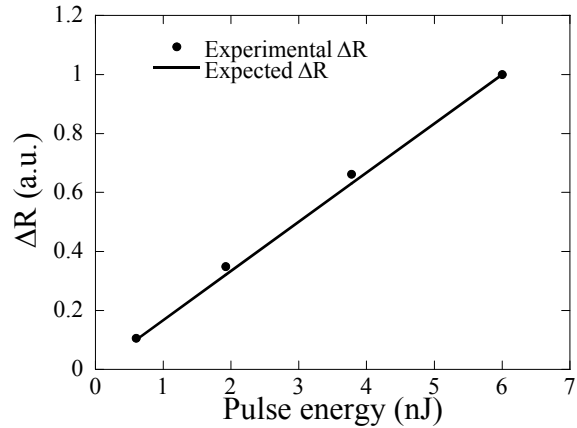

Figure S1: Fluence dependent reflectivity at 2,000 nm probe wavelength shows a linear response of the signal amplitude.

**Supporting Information 2 | Rapid randomization of carriers**

We observed the rise of the transient signal at maximum possible time resolution permitted by our system for armchair pumping and zigzag pumping keeping the probe polarization fixed along the armchair direction and found that the transient signal begins and peaks at the same delay for both cases. This indicates that carriers generated in the zigzag direction rapidly randomize their momentum and orient in the armchair direction. This process is faster than our measurement time resolution as shown in Figure S2.

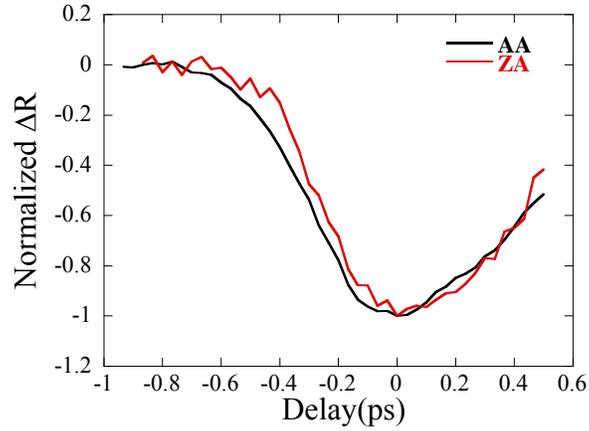

Figure S2: Fast randomization in k-space is observed by probing along the armchair direction but pumping along two perpendicular directions. The ZA (zigzag pump, armchair probe) case is noisier due to lower absorption leading to reduced signal-to-noise. AA stands for armchair pump and armchair probe.

**Supporting Information 3 | Transmittance of bulk black phosphorus**

A piece of black phosphorus was exfoliated to entirely cover the scotch tape and form a uniform layer. Power transmitted through the tape with flakes ($P_{flake}$), through the tape without flakes ($P_{tape}$) and through air ($P_{air}$) were recorded and the transmittance of black phosphorus was extracted using equation S3-1. Figure S3 shows the observed transmittance.

$$T = \frac{P_{tape} - P_{flake}}{P_{air}} \quad\quad (S3\text{-}1)$$

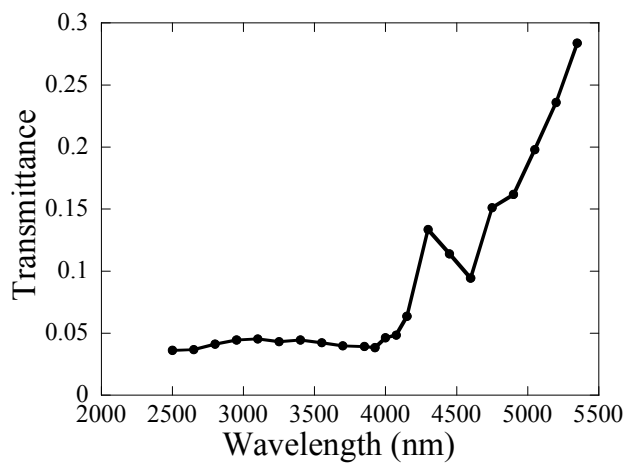

Fig S3: Transmittance of thick layer of exfoliated black phosphorus. The flakes are randomly oriented.

**Supporting Information 4 | Transient data at 5,200 nm**

As the probing wavelength was increased beyond 4,600 nm, the transient signals were completely lost indicating no energy states. An example of data obtained at 5,200 nm is shown in Figure S4.

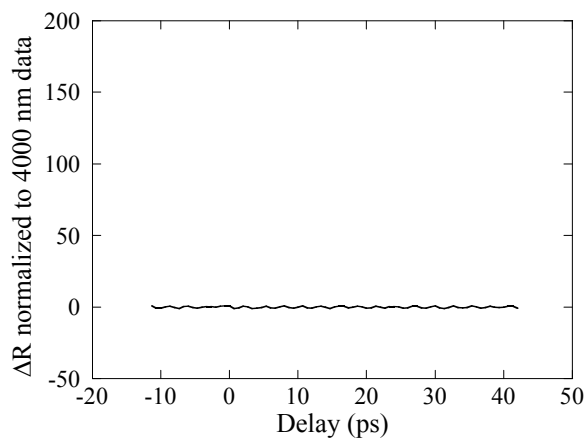

Figure S4: Transient signal at 5,200 nm showing no change in reflectance.

**Supporting Information 5 | Bi-exponential fitting of data**

All of the acquired transient reflectance curves can be fit with the bi-exponential equation S5-1. Example fittings for 2,200 and 4,000 nm probing are shown in Figure S5.

$$\Delta R_{t>0} = A * \exp\left(-\frac{t}{\tau_1}\right) + B * \exp\left(-\frac{t}{\tau_2}\right) + C \tag{S5-1}$$

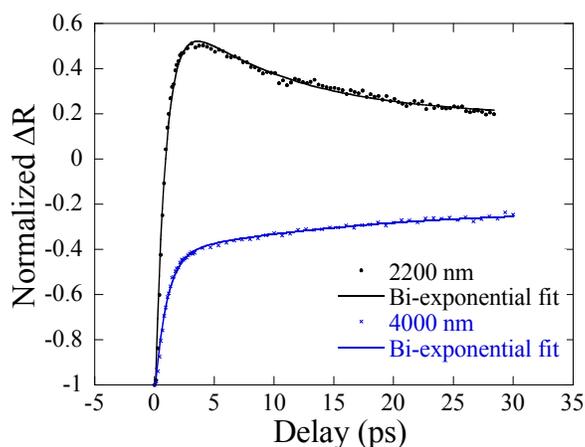

Figure S5: A least squares fitting routine is used to fit a bi-exponential curve for the transient data starting from delay time of 0 picoseconds.

**Supporting Information 6 | Transient data for Zigzag direction probing**

To better understand these anisotropic properties we performed the tests at 2,200 nm and 4,000 nm for all polarization/sample orientation combinations. The transient signals for the four combinations, namely, armchair pump-armchair probe (AA), armchair pump-zigzag probe (AZ), zigzag pump-armchair probe (ZA) and zigzag pump-zigzag probe (ZZ) are shown in Figure S6. All transient data can be fit with the bi-exponential function Eq. S5-1 and the fast relaxation time constant for zigzag probing is found to be about 0.9 ps at 4,000 nm, which is close to the value when using armchair probing.

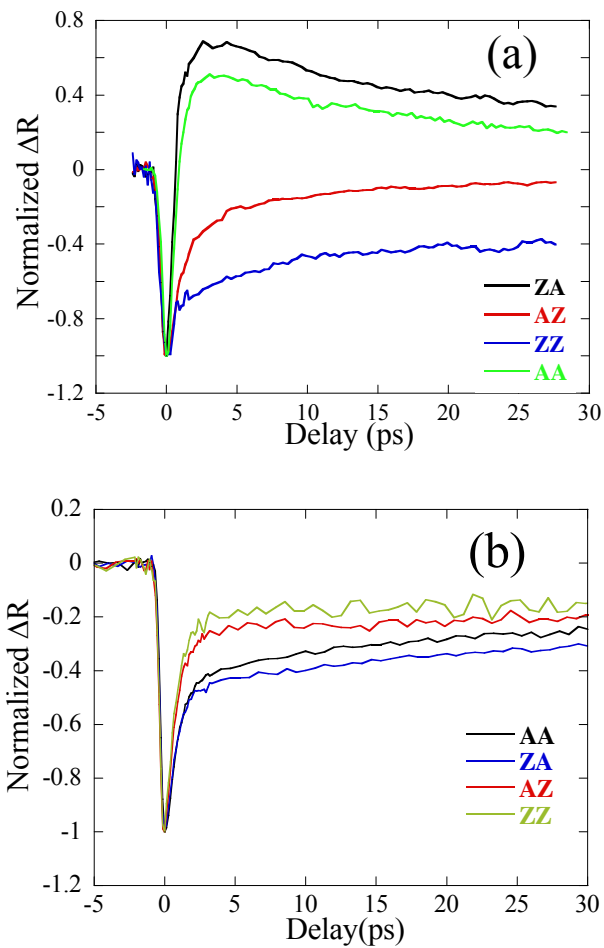

Figure S6: Transient reflectivity for ZA, AZ and ZZ and AA combinations using (a) 2,200 nm and (b) 4,000 nm probe wavelength.

**Supporting Information 7 | The effect of surface doping on the ultrafast reflectivity signals**

Figure S7-1 shows the sample's band diagram with surface doping.

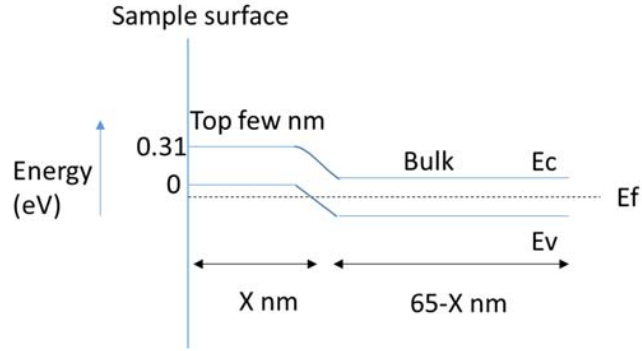

Figure S7-1: Band diagram of surface doped black phosphorus

We can first calculate the hole density N as a function of energy (in the valance band) as follows[1]:

$$N(E) = (1 - f(E))D(E) \tag{S7-1}$$

$$f(E) = \frac{1}{\exp(\frac{E-E_f}{kT})} \tag{S7-2}$$

$$D(E) = \frac{\sqrt{2}(E_h-E)^{\frac{1}{2}}(m_{de})^{3/2}}{\pi^2 \hbar^3} \tag{S7-3}$$

$$m_{de} = (m_1^* m_2^* m_3^*)^{1/3} \tag{S7-4}$$

where $N$ is the carrier density, $f$ is the Fermi-Dirac distribution, $D$ is the density of states, $m_{de}$ is the density of states effective mass, $\hbar$ is the reduced Planck's constant, and $k$ is the Boltzmann constant. The zero reference is taken as the valence band maximum as shown in Figure S7-1. The effective mass values are obtained from literature[2]. The choice of Fermi-level $E_f$ determines the population distribution. For $E_f = -0.077$ eV, which is obtained iteratively to match our experimental data (see below, Figure S7-5), the hole distribution is shown in Figure S7-2.

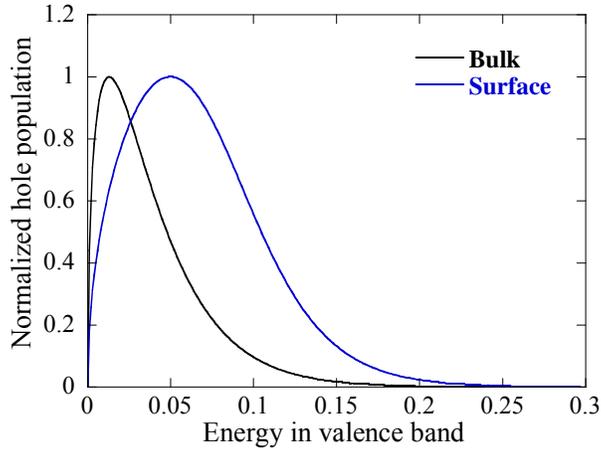

Figure S7-2: Hole population in (a) bulk (b) doped layers

After optical excitation, new carriers are generated which thermalize with the existing population. This can be modeled as a perturbation in the Fermi level. The calculated change in carrier population using Eq. S7-1, after applying the perturbation, as a function of energy in the valence band is shown in Figure S7-3.

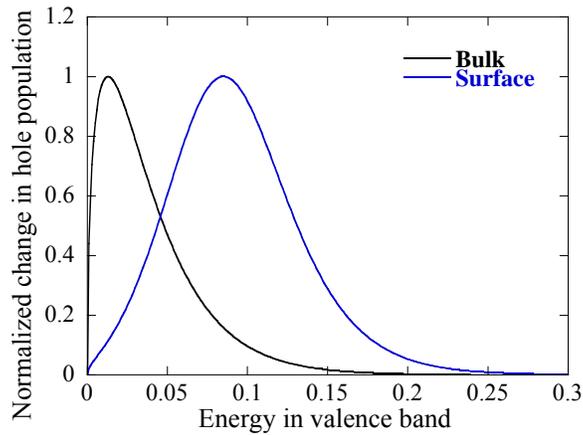

Figure S7-3: Change in hole population for small perturbation in Fermi level

The change in the imaginary part of the dielectric constant, $\Delta\varepsilon_2$, is proportional to the change in carrier population but with an opposite sign, i.e. for an increase in population, $\Delta\varepsilon_2$ decreases. The change in the real part of the dielectric constant, $\Delta\varepsilon_1$, is then obtained by Kramers-Kronig integration. The calculated change in optical constants as a function of energy in the valence band is presented in Figure S7-4.

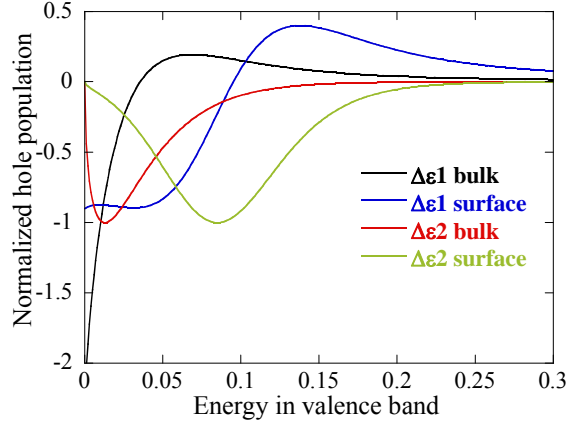

Figure S7-4: Change in dielectric constant after photoexcitation

We can then calculate the change in reflectivity as a result of the change in optical constants using multilayer reflectivity model as[3]:

$$\frac{\Delta R}{R} = \frac{\left(\frac{\partial R}{\partial \varepsilon_1}\Big|_{surf}\Delta\varepsilon_{1,surf}+\frac{\partial R}{\partial \varepsilon_2}\Big|_{surf}\Delta\varepsilon_{2,surf}+\frac{\partial R}{\partial \varepsilon_1}\Big|_{bulk}\Delta\varepsilon_{1,bulk}+\frac{\partial R}{\partial \varepsilon_2}\Big|_{bulk}\Delta\varepsilon_{2,bulk}\right)}{R} \qquad (S7\text{-}5)$$

$$r = \frac{\gamma_0 m_{11}+\gamma_0\gamma_s m_{12}-m_{21}-\gamma_s m_{22}}{\gamma_0 m_{11}+\gamma_0\gamma_s m_{12}+m_{21}+\gamma_s m_{22}} \qquad (S7\text{-}6)$$

$$\gamma_0 = n_0\sqrt{\varepsilon_0\mu_0},\ \gamma_s = n_s\sqrt{\varepsilon_0\mu_0} \qquad (S7\text{-}7)$$

$$R = r * r^* \qquad (S7\text{-}8)$$

$n_0$ and $n_s$ are the refractive indices of air and glass (substrate). $m_{11}$, $m_{12}$, $m_{21}$, and $m_{22}$ are the resultant matrix elements. $\varepsilon_0$, $\mu_0$ are the permittivity and permeability of free space. The reflectance (Eq S7-8) was numerically differentiated with respect to each of the four ε's and then multiplied with the obtained Δε's. $\varepsilon_1$ and $\varepsilon_2$ at 1,550 nm was measured to be 10.5 and 2.15, respectively, along the armchair direction. This value was used for all wavelengths as an estimation. The thickness of the layers, X and 65-X (see Figure S7-1), are embedded in the matrix elements and so is the wavelength.

For a given value of dopant density the change in dielectric constant is calculated by perturbing the Fermi-level as discussed previously. The dopant density also yields a surface layer thickness of X = 5.7 nm by using the measured surface carrier density of $7.5 \times 10^{12}$ cm$^{-2}$ [2]. The reflectivity from the entire structure is then calculated.

Various values of the dopant density were used till the calculated ΔR/R as a function of probe wavelength matches the experimental data for wavelengths around 2,000 nm as shown in Figure S7-5. For the data shown in Figure S7-5, the dopant density value used is $1.3 \times 10^{19}$ cm$^{-3}$ (Fermi-level of -0.077 eV), and the corresponding X is 5.7 nm.

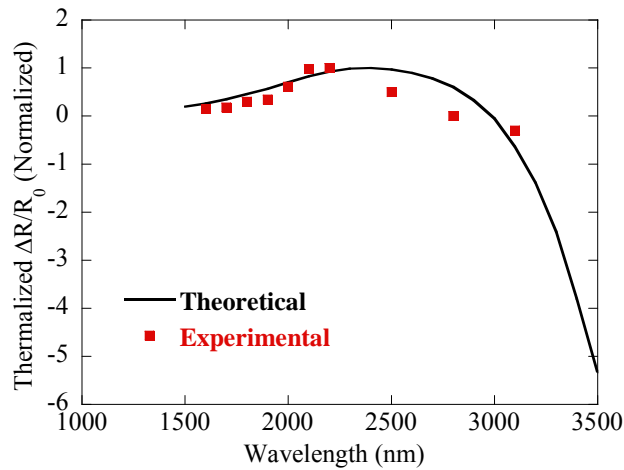

Figure S7-5: Theoretical estimate of ΔR/R and experimental values.

The positive reflectivity does not appear in the data using zigzag direction probing (Figure S6b) because for the same Fermi level, the population density in the zigzag direction is one order of magnitude higher than in the armchair direction owing to the large effective mass. The sensitivity is thus lost in the large background surface hole density. These features in the transient reflectivity signals were also observed in other works[4, 5] at similar wavelengths but no detailed discussions were given.

References for Supporting information: